%% file: main.tex
\documentclass[%
 reprint,
superscriptaddress,
nofootinbib,
 amsmath,amssymb,
 aps,
 prx,
]{revtex4-2}

\usepackage{xcolor} 
\usepackage{graphicx}
\usepackage{listings}
\definecolor{codestring}{rgb}{0.60,0.10,0.10}
\definecolor{codecomment}{rgb}{0.30,0.45,0.30}
\lstdefinestyle{pyconsole}{%
  language=Python,
  basicstyle=\scriptsize\ttfamily,
  stringstyle=\color{codestring},
  commentstyle=\itshape\color{codecomment},
  keywordstyle=\bfseries\color{black},
  showstringspaces=false,
  columns=fullflexible,
  keepspaces=true,
  breaklines=false,
  aboveskip=0.6em, belowskip=0.4em,
}
\usepackage{subfigure}
\usepackage{booktabs} 
\usepackage{hyperref}



\begin{document}

\title{Beyond Stoner--Wohlfarth: Machine-Learning Models and Symbolic Regression of Hard-Magnet Properties}
\author{Samuel J. R. Holt}
\affiliation{Max Planck Institute for the Structure and Dynamics of Matter, Hamburg, Germany}
\affiliation{Center for Free-Electron Laser Science, Hamburg, Germany}
\author{Christina Winkler}
\affiliation{Max Planck Computing and Data Facility, Garching, Germany}
\author{Timoteo Colnaghi}
\affiliation{Max Planck Computing and Data Facility, Garching, Germany}
\author{Martin Lang}
\affiliation{Max Planck Institute for the Structure and Dynamics of Matter, Hamburg, Germany}
\affiliation{Center for Free-Electron Laser Science, Hamburg, Germany}
\author{Swapneel A. Pathak}
\affiliation{Max Planck Institute for the Structure and Dynamics of Matter, Hamburg, Germany}
\affiliation{Center for Free-Electron Laser Science, Hamburg, Germany}
\author{Andrea Petrocchi}
\affiliation{Max Planck Institute for the Structure and Dynamics of Matter, Hamburg, Germany}
\affiliation{Center for Free-Electron Laser Science, Hamburg, Germany}
\author{Michael Adams}
\affiliation{Max Planck Institute for the Structure and Dynamics of Matter, Hamburg, Germany}
\affiliation{Center for Free-Electron Laser Science, Hamburg, Germany}
\author{Thomas Schrefl}
\affiliation{University for Continuing Education, Krems, Austria}
\author{Andreas Marek}
\affiliation{Max Planck Computing and Data Facility, Garching, Germany}
\author{Hans Fangohr}
\affiliation{Max Planck Institute for the Structure and Dynamics of Matter, Hamburg, Germany}
\affiliation{Center for Free-Electron Laser Science, Hamburg, Germany}
\affiliation{University of Southampton, Southampton, United Kingdom}

\date{\today}

\begin{abstract}
Predicting the extrinsic properties from hysteresis loops of a magnetic grain, namely the coercive field, remanent magnetisation, and maximum energy product, from its intrinsic micromagnetic parameters is a central problem in permanent-magnet modelling.
Established analytical models provide useful estimates but often neglect nonuniform magnetisation processes, whereas direct micromagnetic simulations are computationally expensive.
In this work, we train machine-learning models on $12012$ micromagnetic simulations of an idealised cubic grain, spanning broad ranges of the saturation magnetisation, exchange constant, and uniaxial anisotropy constant.
Benchmarked against the analytical models on identical held-out data, the machine-learning models predict all three extrinsic properties with substantially lower errors.
Symbolic regression recovers the Kronm\"uller form of the coercive field, with an effective demagnetising factor that depends on the material, and finds new closed-form expressions for the remanence and maximum energy product.
Each law contains at most two fitted constants yet approaches the accuracy of the machine-learning models.
We also investigate the inverse problem of recovering the intrinsic parameters from the three extrinsic properties.
The saturation magnetisation and anisotropy constant are recovered accurately, whereas the exchange constant is not, because it influences the extrinsic properties only weakly.
The trained models are released through the \texttt{mammos-ai} Python package, enabling thousands of candidate parameter sets to be screened in seconds rather than the hours or days required by direct micromagnetic simulation.
\end{abstract}

\maketitle

\section{Introduction}
Mapping the intrinsic properties of magnets to their extrinsic hysteresis properties is fundamental for advancing technologies critical to energy generation, electric motors, and data storage applications~\cite{gutfleisch2011magnetic,coey2010magnetism}.
Despite this importance, relating intrinsic properties, such as the saturation magnetisation ($M_\mathrm{s}$), exchange constant ($A$), and uniaxial anisotropy constant ($K$), to extrinsic properties such as the coercive field ($H_\mathrm{c}$), remanent magnetisation ($M_\mathrm{r}$), and maximum energy product ($BH_\mathrm{max}$) remains challenging.
The relationship is material- and microstructure-dependent, while its direct evaluation through micromagnetic simulation is computationally expensive.

Traditionally, analytical theories such as the Stoner--Wohlfarth and Kronm\"uller models provide frameworks for estimating extrinsic magnetic properties~\cite{Stoner1948,kronmuller1988analysis}.
However, these analytical approaches often fail to account for realistic complexities found in microstructures, resulting in discrepancies between theoretical predictions and experimental measurements.

Micromagnetic simulations bridge this gap by explicitly incorporating detailed microstructural characteristics into computational models, numerically calculating the magnetic state, thus offering higher accuracy compared to analytical methods~\cite{fischbacher2018micromagnetics}. 
Despite their accuracy, the computational intensity and time demands often limit their practicality for rapid exploration and optimisation of material properties.
To address this issue, Machine Learning (ML) has emerged as a powerful tool within computational materials science, capable of efficiently and accurately creating surrogate models to predict complex material responses~\cite{exl2019magnetic}.

A ML model, however, is accurate but often opaque: it predicts the extrinsic properties without providing an equation that can be inspected or compared with the analytical models.
Symbolic regression addresses this by searching directly for closed-form expressions that fit the data~\cite{schmidt2009distilling,cranmer2023interpretable}.
Applied to micromagnetic simulation data, it can show how far the established analytical forms hold, and suggest corrections where they fail.

In this paper, we introduce a machine-learning surrogate model for an idealised isolated cubic grain structure.
We generate an extensive dataset through thousands of micromagnetic simulations, varying intrinsic properties and capturing corresponding extrinsic outcomes.
We benchmark our machine-learning models against the analytical methods on the same simulation data, distil the learned mapping into closed-form laws by symbolic regression, invert the mapping to recover intrinsic parameters from extrinsic properties, and release the trained models for direct use as part of the \texttt{mammos} software suite in the \texttt{mammos-ai} Python package~\cite{fangohr2026mammos,mammos2026}.

\section{Established analytical models}\label{sec:analytical}
In this paper, we restrict our analysis to the case of hard magnetic materials.
In literature, several analytical models have been developed to relate the intrinsic micromagnetic parameters, the saturation magnetisation ($M_\mathrm{s}$), exchange constant ($A$), and uniaxial anisotropy constant ($K$), to the extrinsic magnetic properties.

The most basic question these models can answer is whether a material is hard at all.
A hard magnet retains its magnetisation once the applied field is removed; a soft magnet loses it.
A useful dimensionless measure is the magnetic hardness parameter~\cite{skomski2016magnetic}
\begin{equation}
  \kappa = \sqrt{\frac{K}{\mu_0 M_\mathrm{s}^2}}.
  \label{eq:kappa}
\end{equation}
For a uniformly magnetised sphere, the magnetisation is stable against vortex nucleation only if the anisotropy energy outweighs the demagnetising energy, whose density for a sphere is $\mu_0 M_\mathrm{s}^2/6$~\cite{brown1963micromagnetics,aharoni2000introduction}.
This gives the criterion
\begin{equation}
  K > \tfrac{1}{6}\,\mu_0 M_\mathrm{s}^2,
  \label{eq:hardness_criterion}
\end{equation}
which is equivalent to $\kappa > 1/\sqrt{6} \approx 0.41$.
Materials that satisfy Eq.~\eqref{eq:hardness_criterion} are expected to behave as hard magnets, with a non-zero remanence.
This gives us an analytical classification rule, which we compare against a data-driven classification in Section~\ref{sec:results-classification}.

Along with the geometry and microstructure, the intrinsic parameters also give rise to characteristic length scales that are useful to consider.

The exchange length ($\ell_{\mathrm{ex}}$) quantifies the length scale where competition between exchange and magnetostatic energies are comparable
\begin{equation}
\ell_{\mathrm{ex}} = \sqrt{\frac{2A}{\mu_{0} M_\mathrm{s}^{2}}},
\label{eq:exchange_length}
\end{equation}
where $\mu_{0}$ is the permeability of free space~\cite{abo2013definition}. 

The domain-wall width ($\delta$) gives the length scale of the competition between exchange and uniaxial anisotropy~\cite{kittel1949domains,coey2010magnetism}
\begin{equation}
\delta = \sqrt{\frac{A}{K}}.
\label{eq:domain_wall_width}
\end{equation}

Analytical equations for the extrinsic properties can also be formulated from the intrinsic parameters.
The Stoner--Wohlfarth model~\cite{Stoner1948} gives an upper limit on the coercive field, commonly referred to as the anisotropy field $H_\mathrm{A}$.
It assumes coherent rotation of the magnetisation at zero temperature, with no demagnetising fields:
\begin{equation}
  H_\mathrm{A}= \frac{2K}{\mu_{0}M_\mathrm{s}}.
\label{eq:Hc_SW}
\end{equation}

A commonly used model for coercivity in realistic microstructures incorporates microstructural imperfections via the parameters $\alpha$ and $N_\mathrm{eff}$, following Kronm\"uller's approach~\cite{kronmuller1988analysis,Kou1994coercive}:
\begin{equation}
H_\mathrm{c}= \alpha H_\mathrm{A}- N_\mathrm{eff}M_\mathrm{s},
\label{eq:Hc_K}
\end{equation}
where $N_\mathrm{eff}$ is the effective demagnetising factor, and $\alpha$ is a microstructure-dependent parameter.
For a cubic system, as is our case, $N_\mathrm{eff}=1/3$.

For a hard magnet with its easy axis aligned with the field, the remanent magnetisation is commonly approximated as~\cite{coey2010magnetism}:
\begin{equation}
M_\mathrm{r} \approx M_\mathrm{s}.
\label{eq:remanent_magnetisation}
\end{equation}

\begin{figure*}[tbp]
    \centering
    \includegraphics[width=0.95\linewidth]{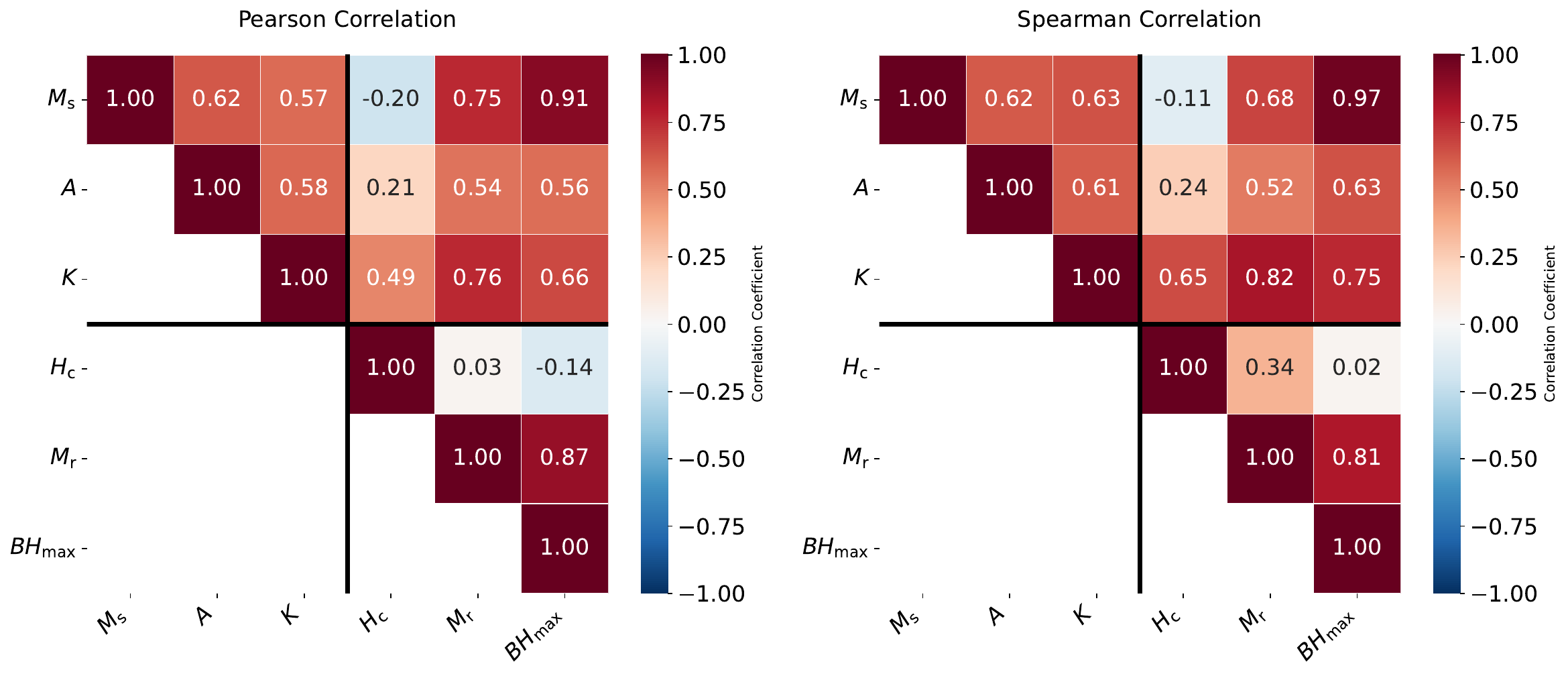}
    \caption{Pearson (left) and Spearman (right) correlation coefficients between intrinsic and extrinsic properties.}
    \label{fig:corr}
\end{figure*}

Lastly, the maximum energy product, an essential figure-of-merit for permanent magnets, can be estimated using~\cite{coey2010magnetism}:
\begin{equation}
BH_\mathrm{max} = \frac{\mu_0M_\mathrm{r}^{2}}{4}.
\label{eq:maximum_energy_product}
\end{equation}

While these analytical relations provide valuable insights, they are idealised and neglect crucial aspects that limit their accuracy in practice~\cite{fischbacher2018micromagnetics}.
Hence, detailed micromagnetic simulations and more sophisticated data-driven modelling approaches are needed.

\section{Methodology}\label{sec:methodology}
For this study, we consider an isolated cubic grain with side length $50\,\mathrm{nm}$ as a benchmark system.
The model includes isotropic exchange, uniaxial anisotropy, Zeeman and demagnetising interactions, while remaining sufficiently simple for comparison with established analytical limiting cases. This allows us to assess whether the machine-learning surrogates recover the known physical trends and accurately capture the deviations.

For data generation, we performed $12012$ micromagnetic simulations using the \texttt{MaMMoS-MuMag} code~\cite{fangohr2026mammos,mammos2026}, a finite-element solver implemented using the \textsc{JAX} library~\cite{jax2018github}.
A single tetrahedral mesh was generated and reused across all simulations.
To ensure that the discretisation resolved the relevant physical length scales, parameter sets were rejected if either the exchange length or the domain-wall width was smaller than the mesh size of $1\,\mathrm{nm}$.

The parameter space consisted of three intrinsic magnetic properties: saturation magnetisation $M_\mathrm{s}$, exchange stiffness $A$, and uniaxial anisotropy constant $K$.The sampled ranges were deliberately broad and were chosen to span a wide micromagnetic parameter space rather than to represent the distribution of a single material family.
The parameter bounds were
\begin{align*}
  7.96 \times 10^{4}\,\mathrm{A/m} 
  &\leq M_\mathrm{s} \leq 
  3.98 \times 10^{6}\,\mathrm{A/m}, \\
  10^{-13}\,\mathrm{J/m} 
  &\leq A \leq 
  10^{-11}\,\mathrm{J/m}, \\
  10^{4}\,\mathrm{J/m^3} 
  &\leq K \leq
  10^{7}\,\mathrm{J/m^3}.
\end{align*}

Each simulation was initialised in a uniformly magnetised state along the $+\hat{z}$ direction.
The uniaxial anisotropy axis was aligned with the same cube edge.
To break symmetry and avoid numerical artefacts during magnetisation reversal, a small misalignment of approximately $1^\circ$ was introduced in the applied magnetic field towards the $+\hat{y}$ direction.
The applied field was then swept quasistatically from $\mu_0 H=+1\,\mathrm{T}$ to $\mu_0 H=-10\,\mathrm{T}$ in steps of $\Delta \mu_0 H = 5\,\mathrm{mT}$.
At each field step, the magnetisation was relaxed to a local energy minimum.
If the net magnetisation reversed before reaching $\mu_0 H=-10\,\mathrm{T}$, the sweep was terminated early to reduce computational cost.
If the magnetisation did not reverse by $\mu_0 H=-10\,\mathrm{T}$ the corresponding outputs were recorded as invalid.
Invalid in this context does not mean that it is physically invalid but that we do not have fully completed micromagnetic simulations for these parameters. 

From each completed hysteresis loop, we extracted the remanent magnetisation $M_\mathrm{r}$, coercive field $H_\mathrm{c}$, and maximum energy product $BH_\mathrm{max}$ using the \texttt{mammos-analysis} package of the \texttt{MaMMoS} software suite~\cite{fangohr2026mammos,mammos2026}.
The maximum energy product was calculated from the demagnetisation curve using a demagnetising coefficient of $1/3$, consistent with the cubic geometry.

For the machine-learning models, the extracted data are prepared as follows.
Samples with $M_\mathrm{r} < 10^{4}\,\mathrm{A/m}$ or $H_\mathrm{c} < 10^{4}\,\mathrm{A/m}$ are excluded before training.
Both the intrinsic inputs and the extrinsic targets are transformed as $x \rightarrow \log(1+x)$, and the inputs are additionally standardised to zero mean and unit variance~\cite{pedregosa2011scikit}.
The targets remain in the transformed space during training.
All predictions are transformed back to physical units for the figures and metrics in the main text; the corresponding transformed-space metrics are given in the Supplemental Material~\cite{supplemental}.
Each model is trained on a random 80\% of the samples and evaluated on the remaining 20\%, with a fixed random seed.
All reported metrics refer to this held-out test set.
The analytical models of Section~\ref{sec:analytical} are evaluated on the identical test samples, with the Kronm\"uller parameter $\alpha$ fitted on the training samples only.

\section{Results}
\subsection{Exploratory analysis of dataset}

We first examine the dataset generated from the micromagnetic simulations.
In total, $12012$ simulations were generated, of which $1624$ produced invalid outputs and were excluded from the valid dataset used for analysis and training.
These invalid outputs correspond to simulations where the magnetisation did not reverse within the applied field range, or where the extrinsic properties could not be extracted.
A further $431$ valid samples fall below the training floors described in Section~\ref{sec:methodology}, leaving $9957$ samples for model training.
The dataset figures in this section show all $10388$ valid samples.

Although $M_\mathrm{s}$, $A$, and $K$ were sampled independently across broad parameter ranges, the final valid dataset is not uniformly distributed across this input space.
This is because the length-scale filtering removes particular regions of parameter space.
For example, parameter sets with low $M_\mathrm{s}$ and high $K$ are more likely to have coercive fields outside the simulated field range, and are therefore more likely to be excluded.
Similarly, the requirement that the exchange length and domain-wall width are larger than the mesh size removes combinations of $A$, $M_\mathrm{s}$, and $K$ that would not be properly resolved.

To quantify relationships between parameters in the final valid dataset, we compute both Pearson~\cite{pearson1895note} and Spearman~\cite{spearman1904proof} correlation coefficients for all pairs of intrinsic and extrinsic properties, as shown in Fig.~\ref{fig:corr}.

The correlations between intrinsic and extrinsic properties highlight several expected physical trends.
The saturation magnetisation $M_\mathrm{s}$ is strongly positively correlated with $BH_\mathrm{max}$, reflecting the importance of remanence for the maximum energy product.
By contrast, $M_\mathrm{s}$ shows a weak negative correlation with $H_\mathrm{c}$, consistent with the reduction of the anisotropy field scale with increasing $M_\mathrm{s}$.
The anisotropy constant $K$ is positively correlated with $H_\mathrm{c}$, as expected from the role of anisotropy in resisting magnetisation reversal.

The exchange stiffness $A$ shows a weaker direct correlation with $H_\mathrm{c}$ than $K$, but it is not completely independent of the extrinsic properties.
This should be interpreted carefully, because part of the apparent correlation of $A$ with $M_\mathrm{r}$ and $BH_\mathrm{max}$ arises from the filtering procedure, which couples the retained values of $A$ to the retained values of $M_\mathrm{s}$ and $K$.
Correlation analysis therefore provides a useful first view of the dataset, but does not by itself establish causal feature importance.

The clear structure in the correlation patterns suggests that a machine-learning model should be able to extract predictive relationships between the intrinsic and extrinsic quantities.
However, the filtering-induced correlations also mean that model performance must be interpreted with care, particularly when assessing the apparent importance of individual intrinsic parameters.

\subsection{Hard and soft magnet regimes}\label{sec:results-classification}
The broad parameter ranges used in this study generate both hard-magnet-like and soft-magnet-like behaviour.
A useful quantity for separating these regimes is the normalised remanence, $M_\mathrm{r}/M_\mathrm{s}$.
Simulations with $M_\mathrm{r}/M_\mathrm{s} \simeq 1$ retain most of their magnetisation after the applied field is removed, and are therefore hard magnets.
By contrast, simulations with $M_\mathrm{r}/M_\mathrm{s} \simeq 0$ lose most of their magnetisation at zero field, and are therefore soft magnets.

Most simulations fall close to one of these two limits, but a small number of intermediate cases require a consistent classification rule.
We therefore use $k$-means clustering~\cite{lloyd1982least} with two clusters, applied to the standardised pair $(M_\mathrm{s},\, M_\mathrm{r}/M_\mathrm{s})$.
This gives a simple, data-driven separation of the simulated hysteresis behaviour without imposing a threshold by hand.
The result does not depend on this choice: clustering on $M_\mathrm{r}/M_\mathrm{s}$ alone produces identical labels, and for data this strongly bimodal the separation is equivalent to a remanence-ratio threshold of about $0.5$.
The resulting clusters are shown in the Supplemental Material.
The clustering separates the valid simulations into $8831$ hard magnet cases and $1557$ soft magnet cases.

\begin{figure}[tbp]
    \centering
    \includegraphics[width=0.95\linewidth]{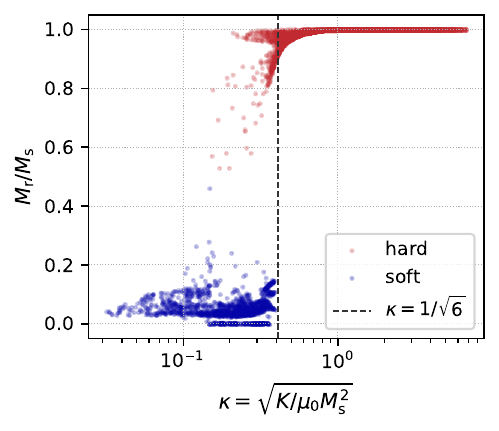}
    \caption{
    Normalised remanence $M_\mathrm{r}/M_\mathrm{s}$ of every valid simulation against the hardness parameter $\kappa$, coloured by the $k$-means classification.
    The dashed line is the analytical hardness criterion $\kappa = 1/\sqrt{6}$ of Eq.~\eqref{eq:hardness_criterion}.
    No soft magnet lies to the right of the line; hard magnets extend below it.
    }
    \label{fig:kappa_classification}
\end{figure}

We can now test the analytical hardness criterion of Eq.~\eqref{eq:hardness_criterion} against this classification.
The comparison is meaningful because the labels are not the output of a competing model: they record the simulated remanence behaviour itself, and the strong bimodality makes them insensitive to how the two groups are separated.
The criterion therefore predicts, from the intrinsic parameters alone, how a grain will behave before it is simulated.
Figure~\ref{fig:kappa_classification} shows the normalised remanence of every valid simulation against the hardness parameter $\kappa$.
The criterion, derived for a uniformly magnetised sphere, describes our cubic grains well: it classifies $96.3\%$ of the samples correctly.
Its errors are entirely one-sided.
No soft magnet is ever classified as hard, while $385$ simulations remain hard below the $\kappa = 1/\sqrt{6}$ boundary; for the hard class, this corresponds to a precision of $1.00$ and a recall of $0.96$.
The sphere-derived criterion is therefore a conservative rule for cubic grains: passing it guarantees hardness in our data, but hard magnets already exist at $\kappa$ values below it.

In the following sections, this separation of clusters is used to distinguish trends that arise from the full simulated parameter space from those that are relevant within the hard-magnet regime.
In the remainder of the paper we consider the hard-magnet subset unless stated otherwise.

\subsection{Analytical models}\label{sec:results-analytical}
\subsubsection{Coercive field \texorpdfstring{$H_\mathrm{c}$}{Hc}}

\begin{figure}[tbp]
    \centering
    \includegraphics[width=0.95\linewidth]{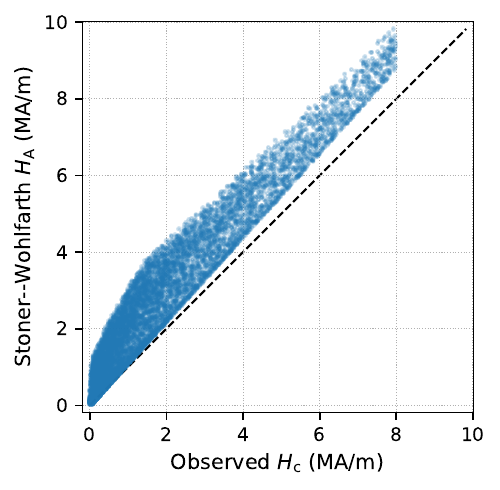}
    \includegraphics[width=0.95\linewidth]{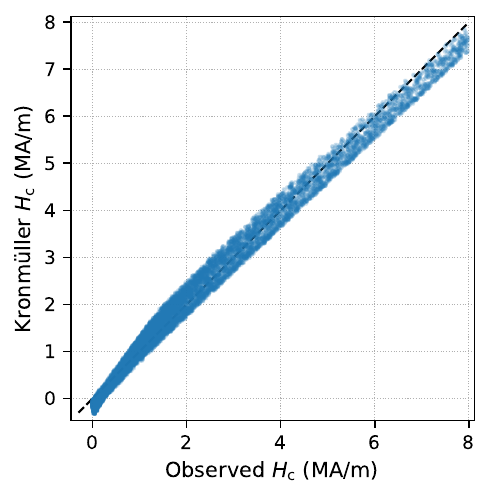}
    \caption{
    Analytical estimates of the coercive field for the hard-magnet-like subset.
    Top: Stoner-Wohlfarth estimate.
    Bottom: fitted Kronm\"uller model.
    The dashed line indicates perfect agreement between the analytical estimate and the simulated value.
    All hard-magnet samples are shown; the errors quoted in the text refer to the held-out test samples only.
    }
    \label{fig:analytical:Hc_models}
\end{figure}
Initially we will consider the forward problem: predicting the extrinsic properties $H_\mathrm{c}$, $M_\mathrm{r}$, and $BH_\mathrm{max}$ from the intrinsic parameters $M_\mathrm{s}$, $A$, and $K$ of a hard magnet.
We first evaluate the analytical coercivity models introduced above on the hard-magnet subset.
As shown in Fig.~\ref{fig:analytical:Hc_models} (top), the Stoner--Wohlfarth estimate of Eq.~\eqref{eq:Hc_SW} systematically overestimates the simulated coercive field.
This is expected, since the model assumes coherent rotation and does not include the nonuniform reversal processes that occur in the simulations.
On the held-out test samples, the Stoner--Wohlfarth estimate misses with an RMSE of $1088\,\mathrm{kA/m}$ (Table~\ref{tab:forward-ml-models}).

The fitted Kronm\"uller model of Eq.~\eqref{eq:Hc_K} performs substantially better, as shown in Fig.~\ref{fig:analytical:Hc_models} (bottom).
Using $N_\mathrm{eff}=1/3$ for the cubic geometry and fitting the reduction parameter to the training samples, we obtain $\alpha=0.849$ and a test RMSE of $195\,\mathrm{kA/m}$ (Table~\ref{tab:forward-ml-models}).
This shows that the coercivity of the hard-magnet subset is overall well described by an anisotropy-field scale corrected by a magnetostatic term.
The model can however fail unphysically: for weakly anisotropic grains the magnetostatic term outweighs the anisotropy term and Eq.~\eqref{eq:Hc_K} predicts a negative coercive field, which occurs for $130$ of the $1757$ test samples.
When computing the metrics, these predictions are clamped to the lowest coercive field in the dataset, $10\,\mathrm{kA/m}$.
The Kronm\"uller model is the strongest analytical benchmark for the machine-learning surrogates.

\subsubsection{Remanent magnetisation \texorpdfstring{$M_\mathrm{r}$}{Mr}}

\begin{figure}[tbp]
    \centering
    \includegraphics[width=0.95\linewidth]{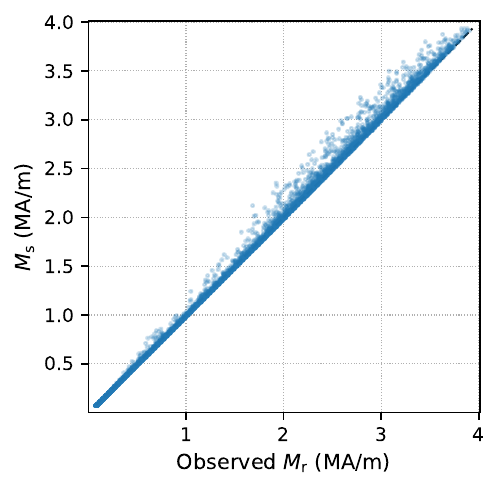}
    \caption{
    Analytical estimate of the remanent magnetisation for the hard-magnet-like subset using $M_\mathrm{r}\approx M_\mathrm{s}$.
    The dashed line indicates perfect agreement between the analytical estimate and the simulated value.
    All hard-magnet samples are shown; the errors quoted in the text refer to the held-out test samples only.
    }
    \label{fig:analytical:Mr_model}
\end{figure}

For the hard-magnet subset, the remanence is simply $M_\mathrm{r}\approx M_\mathrm{s}$, as given by Eq.~\eqref{eq:remanent_magnetisation}.
As shown in Fig.~\ref{fig:analytical:Mr_model}, this approximation agrees closely with the simulations, with a test RMSE of $53\,\mathrm{kA/m}$ (Table~\ref{tab:forward-ml-models}).
Strong agreement is expected here, because the hard-magnet subset was itself selected on $M_\mathrm{r}/M_\mathrm{s}$.
The comparison is therefore a consistency check rather than an independent prediction.
It confirms that hard grains retain almost all of their saturation magnetisation at zero field, and that $M_\mathrm{s}$ is a hard upper limit on $M_\mathrm{r}$.

\subsubsection{Maximum energy product \texorpdfstring{$BH_\mathrm{max}$}{BHmax}}

\begin{figure}[tbp]
    \centering
    \includegraphics[width=0.95\linewidth]{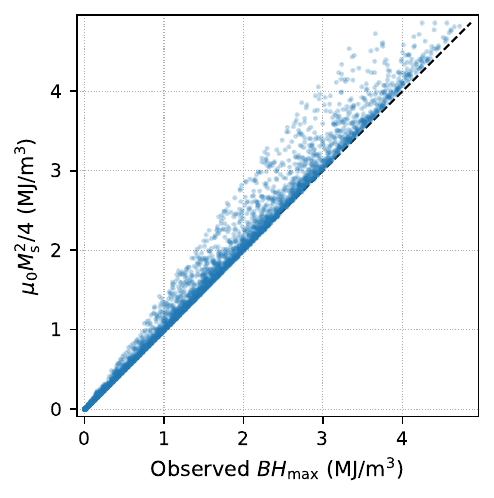}
    \caption{
    Analytical estimate of the maximum energy product for the hard-magnet-like subset.
    The dashed line indicates perfect agreement between the analytical estimate and the simulated value.
    All hard-magnet samples are shown; the errors quoted in the text refer to the held-out test samples only.
    }
    \label{fig:analytical:BHmax_model}
\end{figure}

We next evaluate the analytical estimate for the maximum energy product on the hard-magnet subset.
As shown in Fig.~\ref{fig:analytical:BHmax_model}, the analytical expression of Eq.~\eqref{eq:maximum_energy_product} captures the simulated $BH_\mathrm{max}$ values well, with a test RMSE of $148\,\mathrm{kJ/m^3}$ (Table~\ref{tab:forward-ml-models}).
It remains a systematic upper bound: it assumes an ideal square demagnetisation curve, which the simulated loops do not always reach.

\subsection{Machine Learning models}\label{sec:forward-ml-model}
The accuracy of the analytical models is limited by the simplified physics models and assumptions they are built on.
A machine-learning surrogate carries no such physics based restriction: it learns the relationship between intrinsic and extrinsic properties directly from the simulation data.
We explore five machine-learning models for this task: linear regression and LASSO regression~\cite{tibshirani1996regression,pedregosa2011scikit}, a random forest~\cite{breiman2001random}, a Gaussian process~\cite{rasmussen2006gaussian}, and a fully connected neural network~\cite{paszke2019pytorch}.
The data preparation follows Section~\ref{sec:methodology}, and the hyperparameter grids of all models are given in the Supplemental Material.

\subsubsection{Results}
Table~\ref{tab:forward-ml-models} reports the errors of all five models on the held-out test set, together with the analytical estimates of Section~\ref{sec:results-analytical} evaluated on the identical test samples and the symbolic-regression laws derived in Section~\ref{sec:symbolic-regression}.
The random forest (RF) and the Gaussian process (GP) provide the best predictive performance.
For the coercive field, the fitted Kronm\"uller model reaches a test RMSE of $195\,\mathrm{kA/m}$; the RF reduces this to $92\,\mathrm{kA/m}$ and the GP to $18\,\mathrm{kA/m}$.
For the remanence, the RF improves the analytical estimate from $53$ to $11\,\mathrm{kA/m}$, and for the energy product from $148$ to $17\,\mathrm{kJ/m^3}$.
Every extrinsic property is predicted more accurately by a surrogate than by its best analytical counterpart.

We adopt the RF as the working model in the following.

Although the GP achieves marginally the highest accuracy for every target --- and, within the sampled parameter range verified by the workflow, its predictions constitute genuine interpolation rather than extrapolation --- the accuracy gap between the two models is small. Following the principle of parsimony, we therefore favour the less complex model when predictive performance is comparable.

Several practical considerations support this choice. Screening
10000 candidate materials requires a fraction of a second with the RF, compared to several seconds with the GP. The RF additionally provides feature importances, offering interpretability that the GP lacks. Its per-prediction jackknife uncertainties are of comparable cost to the GP's predictive variance, yet avoid storing a covariance matrix that scales quadratically with the training set size.

Given the negligible difference in accuracy, these advantages in computational efficiency, interpretability, and scalability make the RF the preferred model for the present study.
The fitted weights of the RF models are available on HuggingFace~\cite{mammos_ai_models} as ONNX~\cite{bai2019onnx} files.

\begin{figure*}[tbp]
    \centering
    \subfigure[Coercive Field Train]{\includegraphics[width=0.36\linewidth]{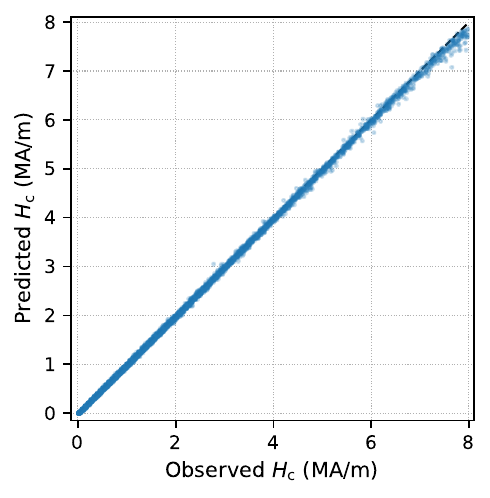}}
    \subfigure[Coercive Field Test]{\includegraphics[width=0.36\linewidth]{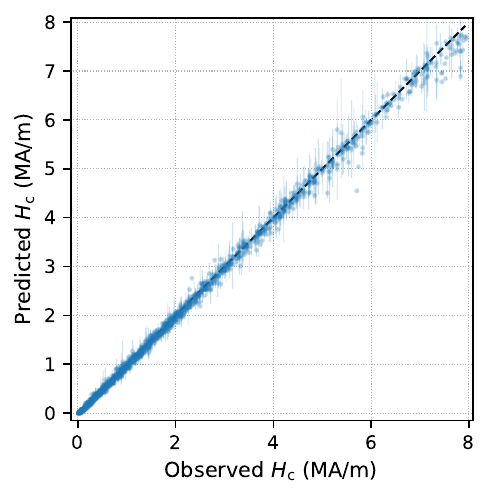}}\\
    \subfigure[Remanent Magnetisation Train]{\includegraphics[width=0.36\linewidth]{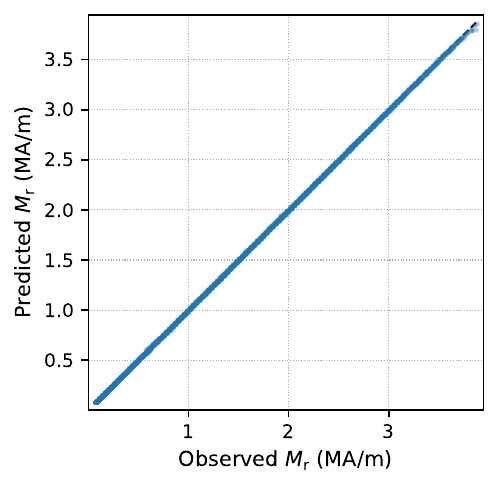}}
    \subfigure[Remanent Magnetisation Test]{\includegraphics[width=0.36\linewidth]{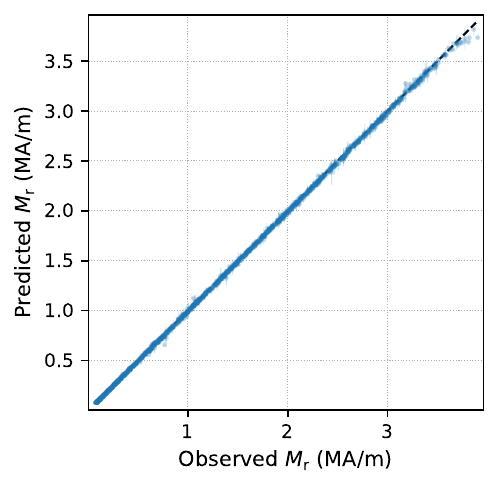}}\\
    \subfigure[Maximum Energy Product Train]{\includegraphics[width=0.36\linewidth]{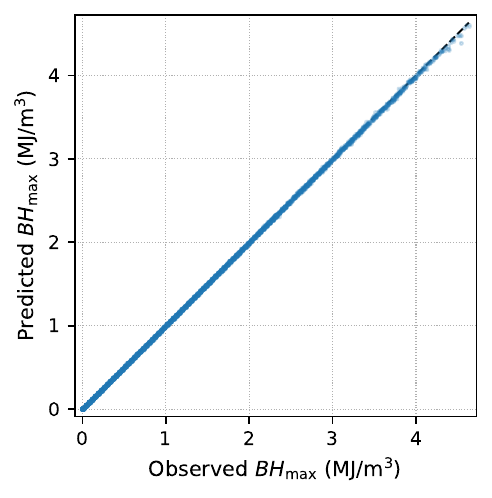}}
    \subfigure[Maximum Energy Product Test]{\includegraphics[width=0.36\linewidth]{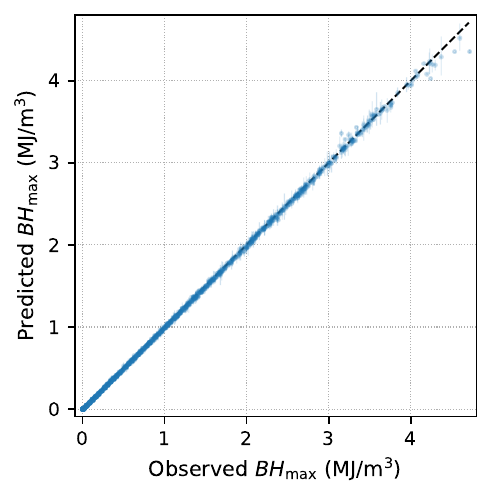}}
    \caption{Random forest predictions of the extrinsic properties against the simulated values for the hard-magnet subset.
    The test panels show the held-out test set with jackknife confidence estimates as error bars.
    The dashed lines indicate perfect agreement.}
    \label{fig:RF_forward}
\end{figure*}

Figure~\ref{fig:RF_forward} compares the RF predictions with the simulated values for all three targets, with the training sets shown in Figs.~\ref{fig:RF_forward}(a), (c), and (e) and the held-out test sets in Figs.~\ref{fig:RF_forward}(b), (d), and (f).
The error bars on the test sets are per-prediction confidence estimates computed with the infinitesimal jackknife method~\cite{Wager2014-qk}.

\begin{table*}[tbp]
\centering
\begin{tabular}{lcccccc}
\toprule
 & \multicolumn{2}{c}{$H_\mathrm{c}$} & \multicolumn{2}{c}{$M_\mathrm{r}$} & \multicolumn{2}{c}{$BH_\mathrm{max}$} \\
\cmidrule(lr){2-3}\cmidrule(lr){4-5}\cmidrule(lr){6-7}
Model & MAPE (\%) & RMSE (kA/m) & MAPE (\%) & RMSE (kA/m) & MAPE (\%) & RMSE (kJ/m$^3$) \\
\midrule
\input{tables/t1_forward_main.tex} 
\bottomrule
\end{tabular}
\caption{Errors of the forward models on the held-out test set, in physical units, for the hard-magnet subset: mean absolute percentage error (MAPE) and root-mean-square error (RMSE). Machine Learning models in the top of the table: Linear Regression (LP), Random Forest (RF), Gaussian Process (GP), LASSO regression (LASSO), Fully Connected Neural Network (FCNN). These are followed by the Symbolic Regression (SR) line, and the analytical models in the bottom of the table. The analytical models are evaluated on the identical test samples; $\alpha$ in the Kronm\"uller model is fitted on the training set only. SR denotes the symbolic-regression laws of Eqs.~\eqref{eq:sr_hc}--\eqref{eq:sr_bhmax}, with constants fitted on the training set only. Full log-space metrics are given in the Supplemental Material.}
\label{tab:forward-ml-models}
\end{table*}

\subsubsection{Feature Importance}
The feature importance of the forward RF model is shown in the Supplemental Material.
It is quantified by the mean decrease in impurity: each node of a regression tree splits on the input variable that most reduces the variance of the target values within that node, and a variable's importance is the total variance reduction it produces, summed over all its splits in every tree and normalised so that the importances of all inputs sum to one.
$M_\mathrm{s}$ carries $0.80$ of the importance and $K$ carries $0.20$, while $A$ contributes $0.002$. Within the sampled parameter space, the extrinsic properties of an ideal hard grain are therefore determined predominantly by
$M_\mathrm{s}$ and $K$.
This is consistent with the simple analytical benchmark expressions used here, Eqs.~\eqref{eq:Hc_SW}--\eqref{eq:maximum_energy_product}, which contain no explicit dependence on $A$. The exchange constant nevertheless enters through the characteristic length scales of Eqs.~\eqref{eq:exchange_length} and \eqref{eq:domain_wall_width}, and can affect the onset of nonuniform magnetisation states and reversal modes. For example, analytical vortex-nucleation models for spherical nanoparticles of radius $R$ contain finite-size exchange corrections proportional to $A/R^2$~\cite{adams2026minimal}.
To confirm the importance of $A$, we retrain the model without $A$ (Supplemental Material).
Removing $A$ degrades every target: the test RMSE grows from $92$ to $128\,\mathrm{kA/m}$ for the coercive field, from $11$ to $14\,\mathrm{kA/m}$ for the remanence, and from $17$ to $23\,\mathrm{kJ/m^3}$ for the maximum energy product.
The influence of $A$ on the extrinsic properties is therefore small, but not negligible, and we keep it as an input. This observation matters for the inverse problem below.
 
\subsection{Symbolic regression}\label{sec:symbolic-regression}
The ML models considered above range in interpretability from linear regression to the black-box neural network.
Symbolic regression offers a middle ground: a data-driven search over closed-form expressions that fit the data.
We use PySR~\cite{cranmer2023interpretable}.

To make the search efficient and its results physically meaningful, we work in dimensionless form.
Each extrinsic property is scaled by its natural analytical unit: the coercive field by the anisotropy field of Eq.~\eqref{eq:Hc_SW}, the remanence by $M_\mathrm{s}$, and the energy product by $\mu_0 M_\mathrm{s}^2/4$.
The dimensionless inputs are the hardness parameter $\kappa$ of Eq.~\eqref{eq:kappa} and the reduced grain size $\tilde{L} = L/\ell_\mathrm{ex}$.
For our fixed geometry these are the only independent dimensionless combinations of $M_\mathrm{s}$, $A$, and $K$, so this representation loses no information.
The search runs on the training samples of the hard-magnet subset only.
From the resulting Pareto front of candidate equations we select the expression with the best trade-off between accuracy and complexity.
We then simplify it by rounding exponents and near-rational constants to exact values wherever this does not degrade the fit on the training set.
The free constants of the final expressions are refitted on the training set by least squares, and the resulting laws are evaluated on the same held-out test set as all other models.

This procedure discovers one equation per extrinsic property:
\begin{align}
  H_\mathrm{c} &= \left[\alpha - n\,\frac{\ln\tilde{L}}{\kappa}\right] H_\mathrm{A},
  \label{eq:sr_hc}\\
  M_\mathrm{r} &= \left(1 - \varepsilon_m \frac{\tilde{L}}{\kappa^4}\right) M_\mathrm{s},
  \label{eq:sr_mr}\\
  BH_\mathrm{max} &= \left(1 - \varepsilon_b \frac{\tilde{L}}{\kappa^4}\right)^{\!2} \frac{\mu_0 M_\mathrm{s}^2}{4},
  \label{eq:sr_bhmax}
\end{align}
with fitted constants $\alpha = 0.942 \pm 0.004$, $n = 0.0921 \pm 0.0004$, $\varepsilon_m = (5.18 \pm 0.02)\times10^{-5}$, and $\varepsilon_b = (7.81 \pm 0.03)\times10^{-5}$.
The quoted uncertainties are the standard errors of the least-squares refit on the training set, obtained from the covariance matrix of the fit; $\alpha$ and $n$ are strongly correlated ($\rho = +0.96$).

\begin{figure*}[tbp]
    \centering
    \subfigure[Coercive Field Train]{\includegraphics[width=0.36\linewidth]{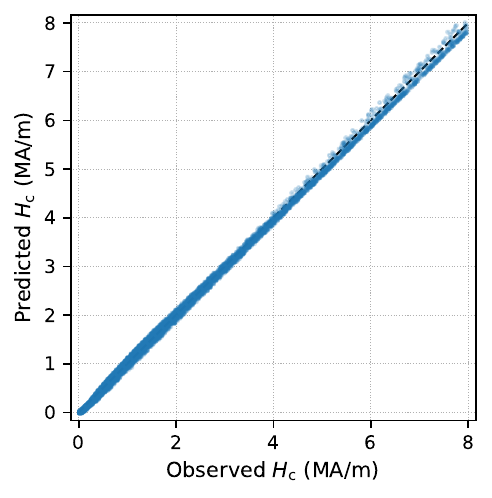}}
    \subfigure[Coercive Field Test]{\includegraphics[width=0.36\linewidth]{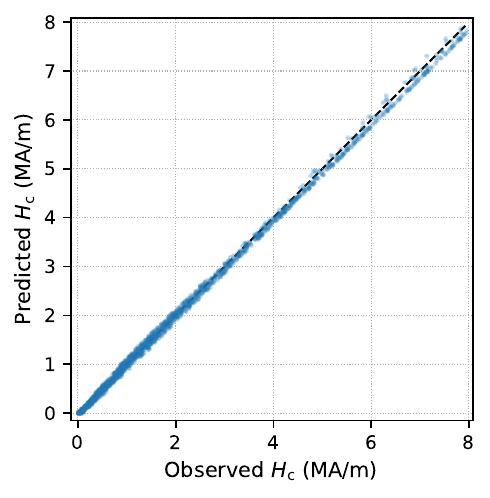}}\\
    \subfigure[Remanent Magnetisation Train]{\includegraphics[width=0.36\linewidth]{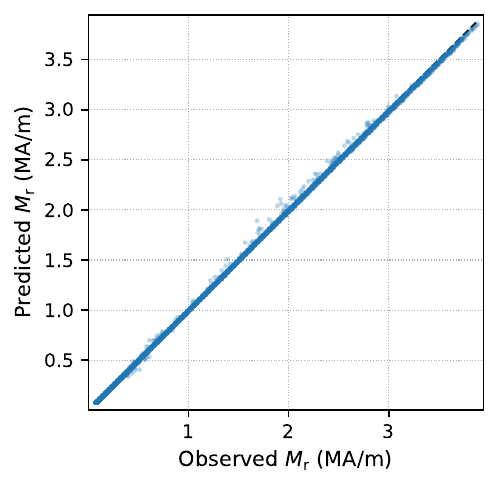}}
    \subfigure[Remanent Magnetisation Test]{\includegraphics[width=0.36\linewidth]{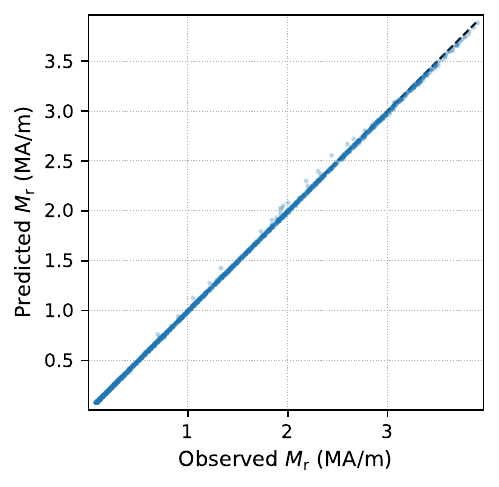}}\\
    \subfigure[Maximum Energy Product Train]{\includegraphics[width=0.36\linewidth]{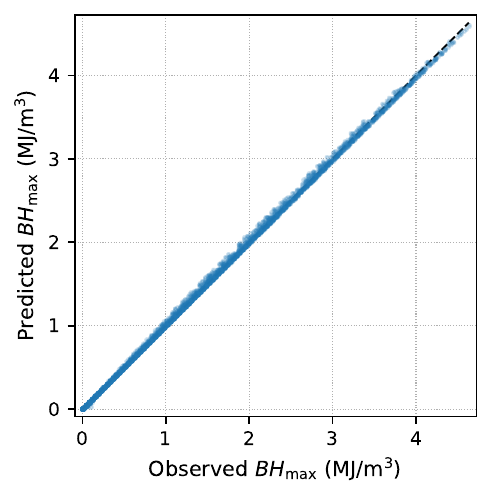}}
    \subfigure[Maximum Energy Product Test]{\includegraphics[width=0.36\linewidth]{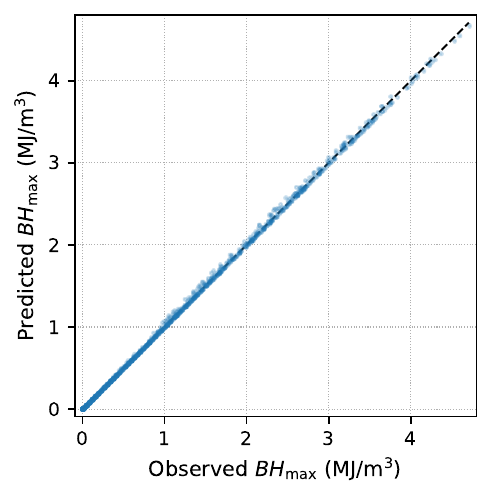}}
    \caption{Symbolic-regression predictions of the extrinsic properties against the simulated values for the hard-magnet subset, using Eqs.~\eqref{eq:sr_hc}--\eqref{eq:sr_bhmax} with constants fitted on the training set only.
    The dashed lines indicate perfect agreement.}
    \label{fig:SR_forward}
\end{figure*}

Each equation contains at most two fitted constants, yet approaches the accuracy of the ML models, and for the coercive field exceeds the RF (Table~\ref{tab:forward-ml-models}).
Only the GP remains clearly more accurate.
Figure~\ref{fig:SR_forward} compares the predictions of the three laws with the simulated values.

The discovered equations are compact and can be interpreted physically.
Multiplying out Eq.~\eqref{eq:sr_hc} gives
\begin{equation}
  H_\mathrm{c} = \alpha H_\mathrm{A} - \left(2n\,\kappa \ln\tilde{L}\right) M_\mathrm{s},
  \label{eq:sr_hc_expanded}
\end{equation}
which is exactly the Kronm\"uller form of Eq.~\eqref{eq:Hc_K}, in which the effective demagnetising factor is no longer a fitted constant but a function of the material, $N_\mathrm{eff} = 2n\,\kappa \ln\tilde{L}$.
A similar logarithmic dependence has been proposed before: Kronm\"uller and F\"ahnle suggested $N_\mathrm{eff} \propto \ln(D/\delta)$ for the grain-size dependence of coercivity~\cite{kronmuller2003micromagnetism}.
Equation~\eqref{eq:sr_mr} is the analytical estimate of Eq.~\eqref{eq:remanent_magnetisation} with a correction that grows when the grain spans many exchange lengths.
This is precisely the regime where the remanent state departs from uniform magnetisation towards the flower state~\cite{schabes1988magnetization}.
Equation~\eqref{eq:sr_bhmax}, in turn, is the square of the remanence law scaled by $\mu_0 M_\mathrm{s}^2/4$: the search independently rediscovers the energy-product relation of Eq.~\eqref{eq:maximum_energy_product}, with the remanence correction folded in.
This is to be expected for a square loop.
To our knowledge, Eqs.~\eqref{eq:sr_mr} and \eqref{eq:sr_bhmax} have no closed-form counterpart in the literature.

Because the cube edge and the demagnetising factor are fixed in our dataset, $\tilde{L}$ varies only through $M_\mathrm{s}$ and $A$.
The power of $L$ in the corrections is therefore an ansatz suggested by the dimensionless formulation rather than a validated size dependence; testing it requires simulations at other grain sizes.

\subsection{Inverse design}
We now consider the inverse problem: predicting the intrinsic parameters $M_\mathrm{s}$, $A$, and $K$ from the extrinsic properties $H_\mathrm{c}$, $M_\mathrm{r}$, and $BH_\mathrm{max}$ of a hard magnet.
The analytical models from literature of Section~\ref{sec:analytical} can be partially inverted: Eq.~\eqref{eq:remanent_magnetisation} returns $M_\mathrm{s}$ from $M_\mathrm{r}$, and Eq.~\eqref{eq:Hc_SW} then gives $K$ from $H_\mathrm{c}$.
No analytical relation, however, involves the exchange constant $A$.
We therefore train the same five models as in Section~\ref{sec:forward-ml-model}, with the same hyperparameter grids, cross-validation strategy, and architectures, with inputs and targets exchanged.
The RF and the GP again perform best (Table~\ref{tab:inverse-ml-models}), and we again focus on the RF.

\subsubsection{Results}
\begin{figure*}[tbp]
    \centering
    \subfigure[Saturation Magnetisation Train]{\includegraphics[width=0.36\linewidth]{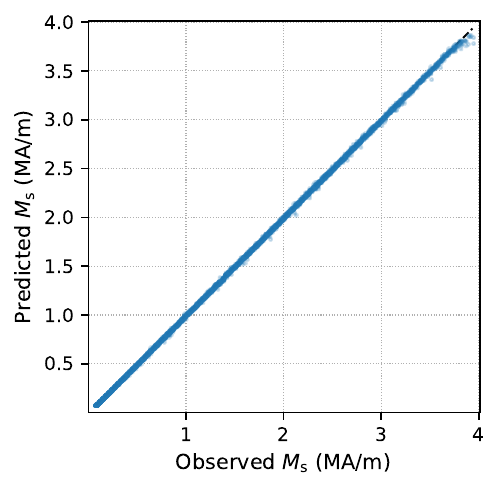}}
    \subfigure[Saturation Magnetisation Test]{\includegraphics[width=0.36\linewidth]{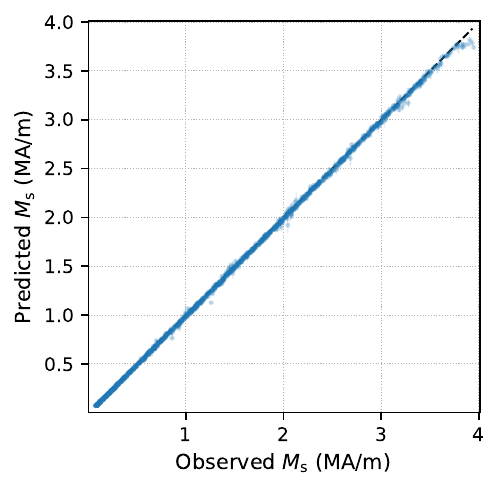}}\\
    \subfigure[Exchange Constant Train]{\includegraphics[width=0.36\linewidth]{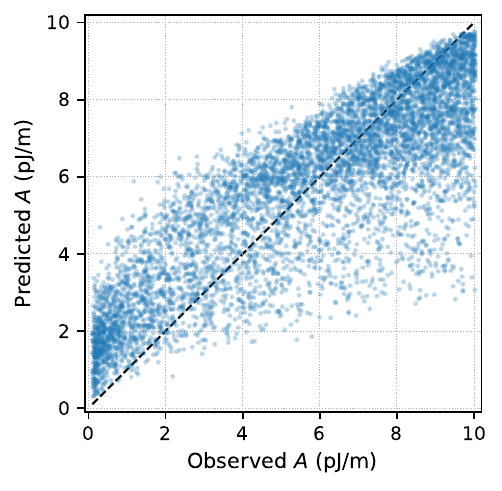}}
    \subfigure[Exchange Constant Test]{\includegraphics[width=0.36\linewidth]{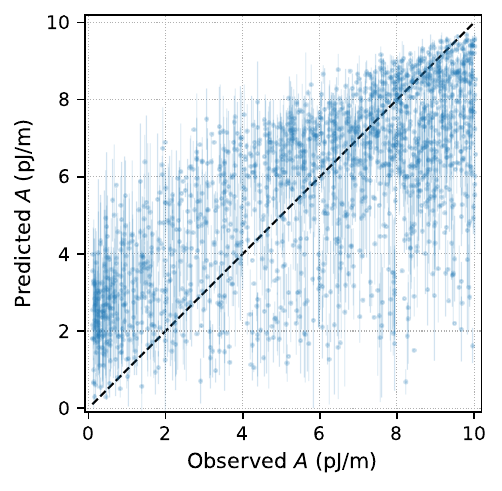}}\\
    \subfigure[Anisotropy Constant Train]{\includegraphics[width=0.36\linewidth]{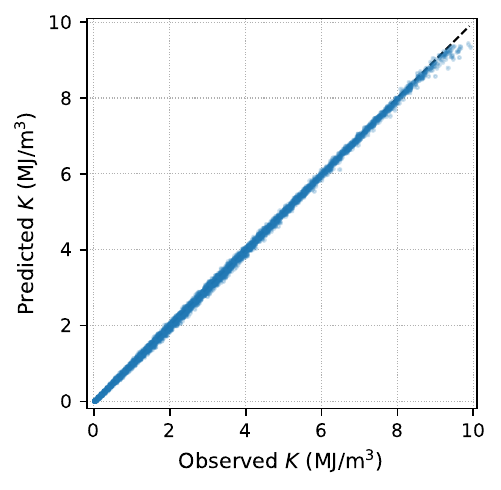}}
    \subfigure[Anisotropy Constant Test]{\includegraphics[width=0.36\linewidth]{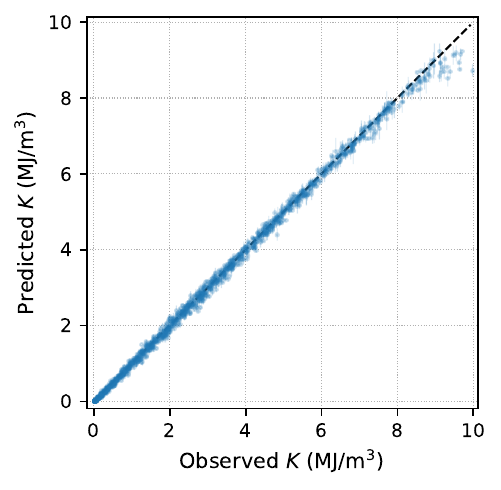}}
    \caption{Random forest predictions of the intrinsic parameters from the extrinsic properties, against the true values for the hard-magnet subset.
    The test panels show the held-out test set with jackknife confidence estimates as error bars.
    The dashed lines indicate perfect agreement.}
    \label{fig:RF_inverse}
\end{figure*}

The three intrinsic parameters are not recovered equally well.
$M_\mathrm{s}$ and $K$ are recovered accurately, with test RMSEs of $14\,\mathrm{kA/m}$ and $100\,\mathrm{kJ/m^3}$, as shown in Figs.~\ref{fig:RF_inverse}(a,b) and (e,f) and Table~\ref{tab:inverse-ml-models}.
The exchange constant $A$ is not: its test RMSE of $2.1\,\mathrm{pJ/m}$ is more than a fifth of the sampled range of $0.1$--$10\,\mathrm{pJ/m}$, and the predictions in Fig.~\ref{fig:RF_inverse}(c,d) scatter far from the diagonal.

This should not be treated as a failure of the regression models, but as evidence that the inverse problem is not uniquely constrained by the three extrinsic properties.
Different values of $A$ produce nearly identical hysteresis loops, so $H_\mathrm{c}$, $M_\mathrm{r}$, and $BH_\mathrm{max}$ are far less sensitive to $A$ than to the other intrinsic parameters.
The feature importance of the forward model (Supplemental Material) supports this from the opposite direction: $A$ only weakly influences the extrinsic properties, so no model can recover it from them.
The symbolic-regression laws make the limit explicit: Eqs.~\eqref{eq:sr_hc}--\eqref{eq:sr_bhmax} invert in closed form for $M_\mathrm{s}$ and $K$, whereas $A$ enters only through $\ln\tilde{L}$ with $\partial\ln H_\mathrm{c}/\partial\ln A \approx 0.04$, so recovering $A$ to $10\%$ would require $H_\mathrm{c}$ to be known to $0.4\%$.

\begin{table*}[tbp]
\centering
\begin{tabular}{lcccccc}
\toprule
 & \multicolumn{2}{c}{$M_\mathrm{s}$} & \multicolumn{2}{c}{$A$} & \multicolumn{2}{c}{$K$} \\
\cmidrule(lr){2-3}\cmidrule(lr){4-5}\cmidrule(lr){6-7}
Model & MAPE (\%) & RMSE (kA/m) & MAPE (\%) & RMSE (pJ/m) & MAPE (\%) & RMSE (kJ/m$^3$) \\
\midrule
\input{tables/t2_inverse_main.tex} 
\bottomrule
\end{tabular}
\caption{Errors of the inverse models on the held-out test set, in physical units, for the hard-magnet subset: mean absolute percentage error (MAPE) and root-mean-square error (RMSE). The analytical models of Section~\ref{sec:analytical} contain no dependence on $A$ and are therefore not included. Full log-space metrics are given in the Supplemental Material.}
\label{tab:inverse-ml-models}
\end{table*}

\subsubsection{Feature Importance}

The feature importance of the inverse RF model is shown in the Supplemental Material.
$BH_\mathrm{max}$ carries most of the importance ($0.74$), followed by $H_\mathrm{c}$ ($0.21$) and $M_\mathrm{r}$ ($0.06$).
The dominance of $BH_\mathrm{max}$ is consistent with the correlation analysis in Fig.~\ref{fig:corr}, which identifies it as the extrinsic quantity most strongly related to the intrinsic parameters.
The ordering of $H_\mathrm{c}$ and $M_\mathrm{r}$, however, differs from their individual correlations: the model exploits their combined information, not each variable alone.

\subsection{Inference on real materials}\label{sec:results-real-materials}

\begin{figure}[tbp]
    \centering
    \includegraphics[width=0.7\linewidth]{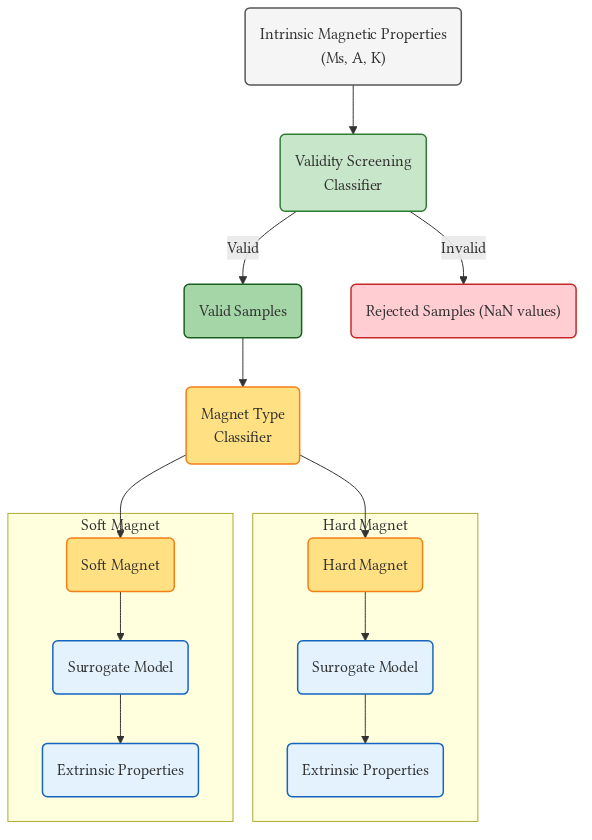}
    \caption{Overview of the ML model pipeline. An invalid sample in Sec.~\ref{sec:methodology}. Two separate surrogate models are trained, one for hard and one for soft magnetic materials. At inference time, the classifier assigns a material type label (hard, soft) based on which the surrogate model is selected.}
    \label{fig:ml_workflow_overview}
\end{figure}

The surrogate forward models of the previous sections can be combined into a single tool for predicting the properties of new materials.
Figure~\ref{fig:ml_workflow_overview} gives an overview of the full pipeline.
A random forest classifier first screens input configurations for validity, since some parameter combinations are outside of the training ranges.
A second classifier, trained on the $k$-means labels described above, then assigns each valid input to the hard or soft regime.
Separate regressors are trained for each regime, with the hard model discussed in detail in the previous sections.
At inference time, a new material is screened, classified, and passed to the regressor of the matching regime.
An end-to-end validation of this chain, in which every simulated parameter set is passed through the released pipeline, is given in the Supplemental Material.

This pipeline, comprising the validity screening, the hard/soft classification, and the trained regressors, is released through the \texttt{mammos-ai} Python package~\cite{fangohr2026mammos,mammos2026}.
We use \texttt{mammos-entity} to define the intrinsic parameters with a value, a unit, and an ontology label, which allows interfaces to be defined to high data-exchange standards. This is not required: the functions also accept plain quantities or values, such as \texttt{numpy} arrays.

\noindent\begin{minipage}{\columnwidth}
Predicting the extrinsic properties of a material takes two lines:
\begin{lstlisting}[style=pyconsole]
>>> import mammos_ai
>>> import mammos_entity as me

>>> # Nd2Fe14B
>>> Ms = me.Entity("SaturationMagnetization", 1.28, "MA/m")
>>> A = me.Entity("ExchangeStiffnessConstant", 7.7, "pJ/m")
>>> K = me.Entity("UniaxialAnisotropyConstant", 4.9, "MJ/m^3")

>>> mammos_ai.Hc_Mr_BHmax_from_Ms_A_K(Ms, A, K)
ExtrinsicProperties(
    Hc=CoercivityHcExternal(4668960.5, A / m),
    Mr=Remanence(1280074.2, A / m),
    BHmax=MaximumEnergyProduct(514737.8, J / m3),
)
\end{lstlisting}
\end{minipage}

Table~\ref{tab:real-materials} shows the released RF pipeline and the symbolic-regression laws of Eqs.~\eqref{eq:sr_hc}--\eqref{eq:sr_bhmax} applied to published micromagnetic parameter sets of well-known hard magnetic materials.
For the four selected materials, dedicated micromagnetic simulations of the same parameter sets agree with the RF predictions to within $3\%$ for the remanence, $5\%$ for the energy product, and $12\%$ for the coercive field; the symbolic-regression laws agree to within $0.6\%$, $1.2\%$, and $6\%$, respectively.

This is expected: both the RF surrogate and the symbolic-regression laws reproduce the micromagnetic simulation of an ideal, defect-free, isolated grain, whereas real magnets reverse at much lower fields because of microstructural defects, misaligned grains, and thermal activation~\cite{fischbacher2018micromagnetics,coey2010magnetism}.

The computational cost of inference is negligible compared with the simulations.
Each micromagnetic simulation behind the training data requires a quasistatic sweep of up to $2200$ field steps, with an energy minimisation at every step, and takes a median of $147$ minutes (mean $211$ minutes) on an NVIDIA V100 GPU.
The released pipeline predicts all three extrinsic properties in $1.5\,\mathrm{s}$ on a single consumer CPU core (AMD Ryzen 7 PRO 7840U), including the validity and regime classifiers and all model-loading overhead, a speed-up of more than three orders of magnitude.
Batched evaluation is faster still: $10000$ parameter sets evaluate in $2.7\,\mathrm{s}$, about $270\,\mathrm{\mu s}$ per material.

\begin{table*}[tbp]
\centering
\begin{tabular}{lcccccccccccc}
\toprule
 & $M_\mathrm{s}$ & $A$ & $K$ & \multicolumn{3}{c}{$H_\mathrm{c}$ (MA/m)} & \multicolumn{3}{c}{$M_\mathrm{r}$ (MA/m)} & \multicolumn{3}{c}{$BH_\mathrm{max}$ (kJ/m$^3$)} \\
\cmidrule(lr){5-7}\cmidrule(lr){8-10}\cmidrule(lr){11-13}
Material & (MA/m) & (pJ/m) & (MJ/m$^3$) & sim. & RF & SR & sim. & RF & SR & sim. & RF & SR \\
\midrule
\input{tables/t3_real_materials.tex} 
\bottomrule
\end{tabular}
\caption{The released RF pipeline and the symbolic-regression laws of Eqs.~\eqref{eq:sr_hc}--\eqref{eq:sr_bhmax} applied to published micromagnetic parameter sets of known hard magnets; each $(M_\mathrm{s}, A, K)$ triple appears together in the cited study.}
\label{tab:real-materials}
\end{table*}

\section{Discussion and conclusions}
Predicting the extrinsic properties of a hard magnet --- the coercive field, remanent magnetisation, and maximum energy product --- from its intrinsic micromagnetic parameters is a central problem in permanent-magnet modelling.
Benchmarked against the analytical models on identical held-out data, machine-learning surrogates trained on approximately $12000$ micromagnetic simulations of an ideal grain predict all three extrinsic properties with substantially lower errors.
Symbolic regression approaches this accuracy with closed-form laws containing at most two fitted constants each, recovering the Kronm\"uller form of the coercive field and, for that property, exceeding the random forest.
The full workflow is a chain of models: a first classifier screens a candidate against the training domain, a second classifier decides whether it is a hard magnet, and a regressor then predicts its extrinsic properties.

We compared five families of ML models as well as symbolic regression for predicting the extrinsic properties from the intrinsic parameters.
Random forests and Gaussian processes perform best, improving substantially on the analytical equations.
Symbolic regression rediscovers the Kronm\"uller form of the coercive field, with a material-dependent demagnetising factor in place of a fitted constant, and finds closed-form corrections to the $M_\mathrm{r}$ and $BH_\mathrm{max}$ estimates that, to our knowledge, are new.
Each law contains at most two fitted constants, yet approaches the accuracy of the surrogates.

For an ideal hard grain, the extrinsic properties are set almost entirely by $M_\mathrm{s}$ and $K$, and the inverse models recover both accurately.
The exchange constant $A$ is different: it shapes the extrinsic properties only weakly, so no model can accurately recover it from $H_\mathrm{c}$, $M_\mathrm{r}$, and $BH_\mathrm{max}$.

This work considers an idealised $50\,\mathrm{nm}$ cubic grain with the applied field aligned to the easy axis.
Grain boundaries, defects, misorientation, polycrystallinity, thermal activation, and chemical disorder, all present in real materials and measurements, are absent.
Microstructure-aware surrogates, trained on simulations of realistic grain ensembles, would extend the approach towards this regime.

The complete ML prediction chain, including the RF classifiers and regressors, is released through the \texttt{mammos-ai} Python package~\cite{fangohr2026mammos,mammos2026}.
The laws of Eqs.~\eqref{eq:sr_hc}--\eqref{eq:sr_bhmax} can be evaluated directly.
A candidate material can be evaluated in seconds, compared with hours for the corresponding simulation.

\section*{Acknowledgments}
This project has received funding from the European Union's Horizon Europe research and innovation programme under the Marie Sk\l{}odowska-Curie grant agreement No.~101152613 and the MaMMoS project, grant agreement No.~101135546.

The authors declare no competing financial interests.

\section*{Data availability}
The simulation dataset and training code are available in the single-grain easy-axis model folder of the MaMMoS ML-models repository~\cite{mammos_mlmodels}.
The trained model weights are published on HuggingFace~\cite{mammos_ai_models} and are usable directly through the \texttt{mammos-ai} package of the MaMMoS software suite~\cite{fangohr2026mammos,mammos2026}.
All results in this paper derive deterministically from these released data, with fixed random seeds throughout.
Every figure in this paper and the Supplemental Material can be reproduced, one script per figure, from the data and code in a dedicated repository~\cite{bsw_figures_repo}.

\bibliography{ref}

\end{document}

%% file: tables/t1_forward_main.tex
LR & 29.8 & 950.4 & 1.0 & 35.1 & 3.6 & 89.3 \\
RF & 5.5 & 92.2 & 0.6 & 11.3 & 1.2 & 16.9 \\
GP & 1.4 & 17.9 & 0.1 & 4.2 & 0.1 & 2.0 \\
LASSO & 30.3 & 807.6 & 1.4 & 38.6 & 3.8 & 89.2 \\
FCNN & 4.3 & 103.4 & 2.4 & 36.3 & 2.4 & 24.6 \\
\midrule
SR & 6.3 & 63.7 & 0.2 & 10.7 & 0.7 & 18.4 \\
\midrule
Stoner--Wohlfarth & 125.4 & 1088.2 & -- & -- & -- & -- \\
Kronm\"uller (fitted $\alpha$) & 19.0 & 195.3 & -- & -- & -- & -- \\
$M_\mathrm{r} = M_\mathrm{s}$ & -- & -- & 1.0 & 52.8 & -- & -- \\
$\mu_0 M_\mathrm{r}^2/4$ & -- & -- & -- & -- & 3.7 & 148.3 \\

%% file: tables/t2_inverse_main.tex
LR & 0.6 & 23.4 & 99.0 & 2.1 & 11.7 & 278.5 \\
RF & 0.7 & 13.9 & 106.0 & 2.1 & 5.5 & 99.9 \\
GP & 0.5 & 34.6 & 105.7 & 2.1 & 5.5 & 147.1 \\
LASSO & 1.2 & 45.1 & 231.0 & 3.0 & 18.7 & 376.5 \\
FCNN & 5.3 & 108.5 & $3.5\times10^{11}$ & $7.8\times10^{9}$ & 5.8 & 125.1 \\

%% file: tables/t3_real_materials.tex
Nd$_2$Fe$_{14}$B~\cite{zickler2015nanoanalytical} & 1.28 & 7.7 & 4.9 & 4.75 & 4.67 & 4.68 & 1.28 & 1.28 & 1.28 & 514 & 515 & 515 \\
$\tau$-MnAl~\cite{arapan2019influence} & 0.64 & 7.6 & 0.7 & 1.43 & 1.42 & 1.34 & 0.64 & 0.63 & 0.64 & 128 & 123 & 127 \\
$\alpha$-MnBi~\cite{tang2022engineering} & 0.67 & 1.1 & 0.9 & 1.51 & 1.69 & 1.54 & 0.67 & 0.66 & 0.67 & 141 & 138 & 140 \\
L1$_0$-FePd~\cite{naganuma2022unveiling} & 0.95 & 6.0 & 2.0 & 2.57 & 2.45 & 2.51 & 0.95 & 0.96 & 0.95 & 283 & 289 & 286 \\